\documentclass{webofc}
\usepackage[varg]{txfonts}   
\usepackage{slashed}
\begin{document}
\title{Confining density functional approach for color superconducting quark matter and mesonic correlations}
%
%
\author{
\firstname{Oleksii} \lastname{Ivanytskyi}
\inst{1}\fnsep\thanks{\email{oleksii.ivanytskyi@uwr.edu.pl}} 
\and
\firstname{David} \lastname{Blaschke}
\inst{1,2,3}\fnsep\thanks{\email{david.blaschke@uwr.edu.pl}} 
\and
\firstname{Konstantin} \lastname{Maslov}
\inst{2}\fnsep\thanks{\email{maslov@theor.jinr.ru}}}
\institute{
Institute of Theoretical Physics, University of Wroclaw, Max Born Pl. 9, 50-204 Wroclaw, Poland 
\and
Bogoliubov Laboratory of Theoretical Physics, JINR Dubna, Joliot-Curie Str. 6, 141980 Dubna, Russia
\and
National Research Nuclear University (MEPhI), Kashirskoe Shosse 31, 115409 Moscow, Russia}
\abstract{
We present a novel relativistic density-functional approach to modeling quark matter with a mechanism to mimic confinement. The quasiparticle treatment of quarks provides their suppression due to large quark selfenergy already at the mean-field level. We demonstrate that our approach is equivalent to a chiral quark model with medium-dependent couplings. The dynamical restoration of the chiral symmetry is ensured by construction of the density functional. Beyond the mean field, quark correlations in the pseudoscalar channel are described within the Gaussian approximation. 
This explicitly introduces pionic states into the model. Their contribution to the thermodynamic potential is analyzed within the Beth–Uhlenbeck framework. The modification of the meson mass spectrum in the vicinity of thee (de)confinement transition is interpreted as the Mott transition. Supplemented with the vector repulsion and diquark pairing the model is applied to construct a hybrid quark-hadron EoS of cold compact-star matter. We study the connection of such a hybrid EoS with the stellar mass-radius relation and tidal deformability. The model results are compared to various observational constraints including the NICER radius measurement of PSR J0740+6620 and the tidal deformability constraint from GW170817. The model is shown to be consistent with the constraints, still allowing for further improvement by adjusting the vector repulsion and diquark pairing couplings.}
\maketitle
\section{Introduction}
\label{sec1}

A principal element of unified description of strongly interacting matter within effective theories is the  hadronization of chiral quark models and the incorporation of a confinement mechanism into them, manifesting the switching between hadronic and quark degrees of freedom. 
The confining aspect of quark matter can be modeled by Cornell-type potentials. 
Within this picture deconfinement of colored degrees of freedom is associated with a decrease of the tension of the string connecting colored constituents and causing a correlation between them. 
Within the string-flip model (SFM), the energy of such a correlation is related to the string length distribution \cite{Horowitz1985,Roepke1986}. 
The average string length is related to the average separation between quarks. 
This separation is connected to the quark number density which, in turn, controls the quark mean-field self-energy. This idea was utilized in Ref.~\cite{Kaltenborn2017} in 
order to develop a density functional approach incorporating the confining aspect of quark matter via the divergence of the quark self-energy at vanishing number density of quarks. 

In this work we do the next step and formulate the SFM-inspired density functional approach in a chirally symmetric way. 
This leads to the appearance of both scalar {\it and} pseudoscalar modes in the model. 
The pseudoscalar mode is most important for the phenomenology of strongly interacting matter in the confined phase since it represents the Goldstone bosons of spontaneously broken chiral symmetry. 
We treat this pseudoscalar mode beyond the mean-field approach within the Gaussian approximation. 
In the two-flavor case this introduces the pion degrees of freedom and in combination with the proposed phenomenological confinement mechanism of quarks in the chirally broken phase it provides a quark-hadron duality within our approach. 

Another important and new aspect of this study is the introduction of the diquark pairing channels into the density functional approach. 
At sufficiently high densities and low temperatures the interaction in these channels causes the formation of a two-flavour color-superconducting phase of quark matter (2SC). We study the thermodynamics this 2SC phase of cold quark matter and derive its equation of state (EoS) within the mean-field approximation.

The recent analysis of observational data on the pulsar PSR J0740+6620 \cite{Riley2021,Miller2021} and the binary neutron star (NS) merger GW170817 \cite{Abbott2018} established challenging constraints on the mass-radius diagram of NS. Fulfilling these constraints requires the EoS of stellar matter to exhibit a significant softening in the density range typical for the intermediate-mass NS with a further stiffening at larger densities typical for NS with masses about two times the solar mass. 
Such a behavior is very unlikely for the scenario of purely hadronic stellar matter since the corresponding EoS consistent with the two solar mass constraint on the NS maximum mass are stiff at all densities. On the other hand, a first order phase transition from hadronic to quark matter can soften the stellar matter EoS in the intermediate density region. 
Therefore, we apply our EoS of cold color superconducting quark matter to model NS with quark cores within the framework of a hybrid quark-hadron EoS.

In the next section, we briefly describe the density functional approach, while sections \ref{sec3} and \ref{sec4} are devoted to the cases of zero chemical potential and zero temperature, respectively. The conclusions are given in section \ref{sec5}.

\section{Density functional approach}
\label{sec2}

We consider the case of two quark flavors labeled by the index $f=u,d$ and represented by the color-flavor-Dirac spinor $q=(u,d)^T$. Their masses and chemical potentials enter the matrices $\hat m={\rm diag}(m_u,m_d)$ and $\hat\mu={\rm diag}(\mu_u,\mu_d)$, respectively. 
The latter includes $\mu_f=\frac{\mu_B}{3}+Q_f\mu_Q$ given 
in terms of the baryon ($\mu_B$) and electric ($\mu_Q$) charge chemical potentials, and the electric charge of quarks $Q_f$. 
Our model accounts for the chirally symmetric interaction in scalar and pseudoscalar channels represented by the density functional $\mathcal{U}$ discussed below, as well as the vector and diquark pairing channels important for the astrophysical applications \cite{Baym2018}. 
The Lagrangian is
\begin{eqnarray}
\label{I}
\mathcal{L}=\overline{q}(i\slashed\partial-\hat m)q-\mathcal{U}-G_V(\overline{q}\gamma_\mu q)^2+G_D
(\overline{q}i\gamma_5\tau_2\lambda_A q^c)(\overline{q}^ci\gamma_5\tau_2\lambda_A q).
\end{eqnarray}
Here $G_V$ and $G_D$ are couplings in vector and diquark channels, respectively, $q^c=i\gamma_2\gamma_0\overline{q}^T$ is the charge conjugate of the quark fields and the color index in the last term is implicitly summed over $A=2,5,7$. Chiral symmetry of $\mathcal{U}$ is provided by its dependence on the argument $(\overline{q}q)^2+(\overline{q}i\gamma_5\vec\tau q)^2$. Following Ref.~\cite{Kaltenborn2017} we expand this 
potential around the mean field expectation values 
$\langle \overline{q}q\rangle$ and $\langle\overline{q}i\gamma_5\vec\tau q\rangle=0$. 
Below all the quantities evaluated at these expectation values are denoted by the subscript index $``MF"$. In this study we use the expansion up to the second order terms, which accounts for the one-loop mesonic correlations of quarks, while the expansion up to the first order terms represents the mean-field approximation. 
Thus,
\begin{eqnarray}\
\label{II}
\mathcal{U}^{(2)}=\mathcal{U}_{MF}+
\left(\overline{q}q-\langle\overline{q}q\rangle\right)\Sigma_{MF}-G_S(\overline{q}q-\langle\overline{q}q\rangle)^2-
G_{PS}(\overline{q}i\gamma_5\vec\tau q)^2.
\end{eqnarray}
Here the mean-field part of the quark self-energy $\Sigma_{MF}$ and the medium-dependent effective couplings 
in the scalar $G_S$ and pseudoscalar $G_{PS}$ channels are defined as
\begin{eqnarray}
\label{III}
\Sigma_{MF}=\frac{\partial\mathcal{U}_{MF}}{\partial\langle\overline{q}q\rangle},\quad
G_S=-\frac{1}{2}
\frac{\partial^2\mathcal{U}_{MF}}{\partial\langle\overline{q}q\rangle^2},\quad
G_{PS}=-\frac{1}{6}
\frac{\partial^2\mathcal{U}_{MF}}{\partial\langle\overline{q}i\gamma_5\vec\tau q\rangle^2}.
\end{eqnarray}

The Lagrangian with the second-order-expanded density functional 
$\mathcal{L}^{(2)}=\mathcal{L}_{\mathcal{U}=0}-\mathcal{U}^{(2)}$ 
is quadratic in all quark bilinears. 
This is equivalent to current-current quark interactions representing a chiral model of the NJL type \cite{Baym2018,Blaschke2014,Klevansky1992,Hufner1994,Zhuang1994,Buballa2005,Ratti2006,Zablocki2010} 
but with medium dependent couplings.

The next step is to bosonize the present model by means of the Hubbard-Stratonovich transformation (see, e.g., Ref. \cite{Blaschke2014} for details). 
This introduces collective fields $\sigma$, $\vec\pi$, $\omega_\mu$ and $\Delta_A$, representing scalar, pseudoscalar, vector and diquark modes coupled to $\overline{q}q-\langle\overline{q}q\rangle$, $\overline{q}i\gamma_5\vec\tau q$, $\overline{q}\gamma_\mu q$ and $\overline{q}i\gamma_5\tau_2\lambda_A q$, respectively. They enter the bosonized Lagrangian as
\begin{eqnarray}
\label{IV}
\hspace*{-.6cm}\mathcal{L}^{bos}+q^+\hat\mu q=\overline{\mathcal{Q}}\mathcal{S}^{-1}\mathcal{Q}-\mathcal{U}_{MF}
+\langle\overline{q}q\rangle\Sigma_{MF}+
\sigma\langle\overline{q}q\rangle-\frac{\sigma^2}{4G_S}-\frac{\vec\pi^2}{4G_{PS}}+
\frac{\omega_\mu\omega^\mu}{4G_V}-\frac{\Delta_A^*\Delta_A^{}}{4G_D}.
\end{eqnarray}
It is written in terms of the Nambu-Gorkov bispinor $\mathcal{Q}^T=\frac{1}{\sqrt{2}}(q~q^c)$ and the propagator
\begin{eqnarray}
\label{IV}
\mathcal{S}^{-1}=\left(
\begin{array}{l}
S^{-1}_+-\sigma-i\gamma_5\vec\tau \cdot\vec\pi\hspace*{1cm}
i\Delta_A^{}\gamma_5\tau_2\lambda_A\\\hspace*{.5cm}
i\Delta_A^*\gamma_5\tau_2\lambda_A\hspace*{1cm}S^{-1}_--\sigma-i\gamma_5\vec\tau^T\cdot\vec\pi
\end{array}\right)
\end{eqnarray}
with $S^{-1}_\pm=i\slashed\partial\pm\slashed\omega-m^*\pm\gamma_0\hat\mu$ and the effective quark mass $m^*_f=m_f+\Sigma_{MF}$. 
The quark fields enter Eq. (\ref{IV}) quadratically and therefore can be integrated out leading to the effective mesonic Lagrangian
\begin{eqnarray}
\label{V}
\mathcal{L}^{eff}=\frac{{\rm Tr}\ln(\beta\mathcal{S}^{-1})}{2\beta V}
-\mathcal{U}_{MF}
+\langle\overline{q}q\rangle\Sigma_{MF}
+\sigma\langle\overline{q}q\rangle+
\frac{\sigma^2}{4G_S}+\frac{\vec\pi^2}{4G_{PS}}-
\frac{\omega_\mu\omega^\mu}{4G_V}+\frac{\Delta_A^*\Delta_A^{}}{4G_D},
\end{eqnarray}
where $\beta=\frac{1}{T}$ is the inverse temperature and 
$V$ is the system volume.

Within our approach the quark confinement is mimicked by a rapid growth of quark self-energy $\Sigma$ in the confining region. 
In agreement with the SFM argument, we assume $\Sigma$ to be inversely proportional to the mean separation between quarks \cite{Horowitz1985,Roepke1986}, i.e. $\Sigma\sim (q^+q)^{-\frac{1}{3}}$ or, equivalently, $\mathcal{U}\sim(q^+q)^{\frac{2}{3}}$. 
The proportionality coefficients in these relations are related to the tension of the confining string. 
Using $\langle\overline{q}i\gamma_5\vec\tau q\rangle=0$ 
and the approximate relation $\langle\overline{q}q\rangle\simeq\langle\overline{q}q\rangle_0+\langle q^+q\rangle$ valid at low $T$ and $\mu$, we parametrize the chirally symmetric interaction potential as
\begin{eqnarray}
\label{VII}
\mathcal{U}=D_0\left[(1+\alpha)\langle \overline{q}q\rangle_0^2
-(\overline{q}q)^2-(\overline{q}i\gamma_5\vec\tau q)^2\right]^{\frac{1}{3}},
\end{eqnarray}
where $\langle \overline{q}q\rangle_0$ is the vacuum value of the chiral condensate and the coupling constant $D_0$ has the meaning of the confining string tension. 
The constant parameter $\alpha$ controls the behavior of the effective quark mass. 
The mean-field self-energy $\Sigma_{MF}$ vanishes for $\alpha\rightarrow\infty$, thus making $\hat m^*=\hat m$. 
For $\alpha=0$ this self-energy diverges for $\langle \overline{q}q\rangle=\langle \overline{q}q\rangle_0$ 
leading to an absolute ``confinement'' of quarks at low $T$ and $\mu$. 
Small positive values of this parameter provide sufficient suppression of quarks in the confining region. 
The adopted parameterization of $\mathcal{U}$ provides a NJL-like expression for the effective mass $\hat m^*=\hat m-2G_{PS}\langle\overline{q}q\rangle$. 
Moreover, replacing the SFM power ''$\frac{1}{3}$'' in Eq. (\ref{VII}) by unity we bring $\mathcal{L}^{(2)}$ to the 
NJL model form \cite{Klevansky1992} up to an insignificant constant term. 
Thus, the present model can be considered as a generalization of the NJL model to the case of density dependent couplings. At the same time $G_{S}\neq G_{PS}$ in the general case due to the expansion of $\mathcal{U}$ around the mean-field solution, which is known to break the chiral symmetry. However, these couplings saturate to the same value $G_\infty=\frac{D_0}{3}(1+\alpha)^{-\frac{2}{3}}\langle \overline{q}q\rangle_0^{-\frac{4}{3}}$, when chiral symmetry gets dynamically restored for vanishing $\langle\overline{q}q\rangle$. 

We fit the vacuum values of the chiral condensate $\langle\overline{q}q\rangle_0=-~2(251~{\rm MeV})^3$, pion mass $M_\pi=140~\rm MeV$, pion decay constant $F_\pi=90~\rm MeV$, and the mean-field value of the pseudocritical temperature $T_c=170~{\rm MeV}$ by choosing the current quark mass $m_u=m_d=m=4.9~\rm MeV$, momentum cutoff $\Lambda=563~\rm MeV$, $\alpha=0.844$, and $D_0\Lambda^{-2}=1.058$. Vector and diquark couplings,  parameterized by $\eta_V\equiv\frac{G_V}{G_\infty}$ and $\eta_D\equiv\frac{G_D}{G_\infty}$, are treated as free parameters. 
Below the pairs of numbers corresponding to these scaled couplings are used in order to label the model parameterizations considered at finite chemical potentials. For example, $(0.2,1.5)$ stands for a model with $\eta_V=0.2$ and $\eta_D=1.5$.

\section{Quark-hadron matter at zero chemical potential}
\label{sec3}

\begin{figure}[t]
\centering
\includegraphics[width=6.4cm,clip]{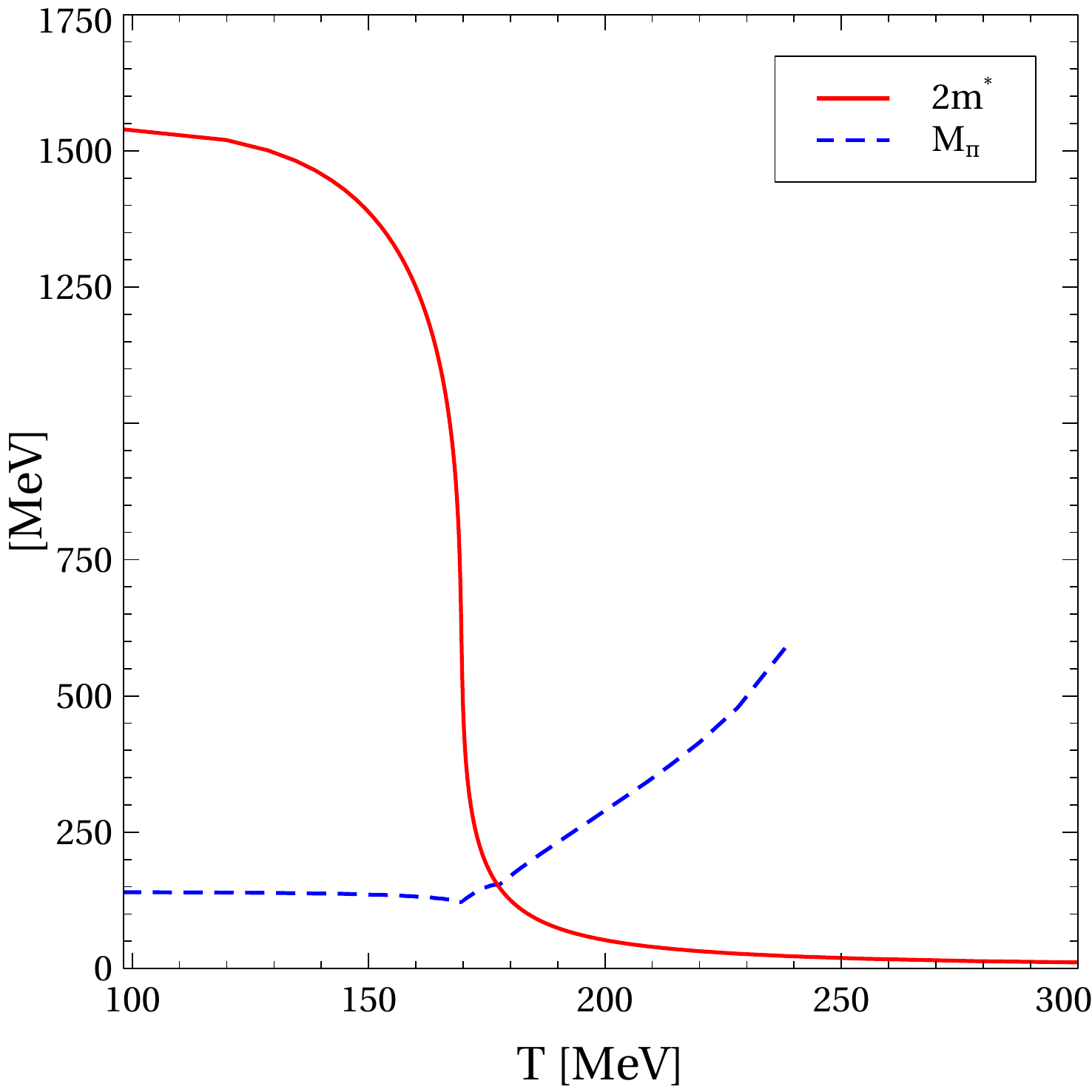}\hspace*{.2cm}
\includegraphics[width=6.4cm,clip]{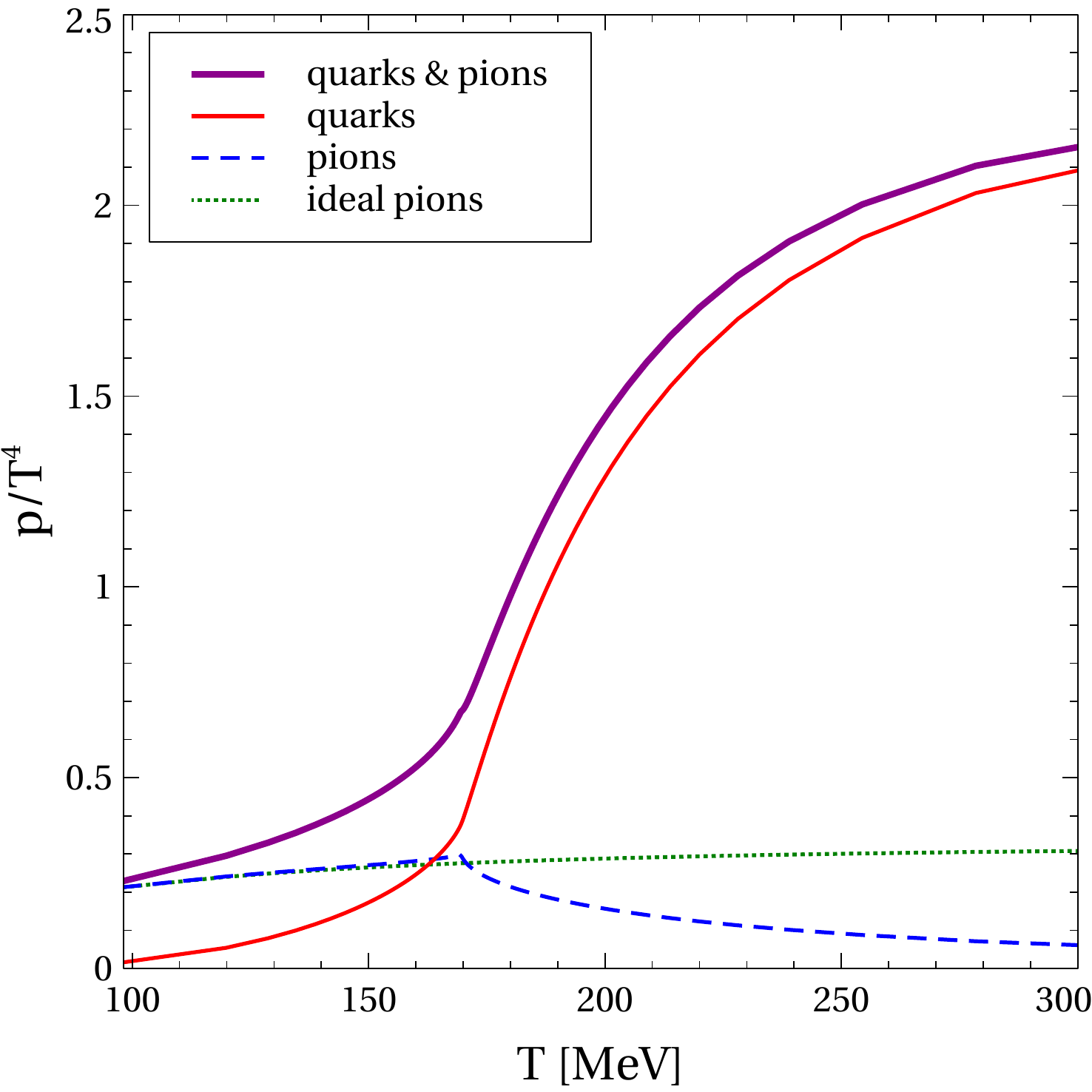}
\caption{{\it Left panel:} temperature dependence of the pion mass $M_\pi$ (blue dashed curve) compared to the two quark decay threshold $2m^*$ (red solid curve). {\it Right panel:} temperature dependence of the scaled pressure of quark-pion system (thick solid magenta curve) compared to partial  contributions of quarks (thin solid red curve), pions (blue dashed curve) and noninteracting pions with mass $M_\pi=140$ MeV (green dotted curve).}
\label{fig1}
\end{figure}

In this regime vector and diquark fields have zero expectation values at the mean-field level. 
On the other hand, quark correlations in the pseudoscalar channel are the most important beyond mean-field. 
We limit the present consideration to these correlations only. 
For this we set $\sigma=\omega_\mu=\Delta_A=0$. 
Note, that $\sigma=0$ within the mean-field approximation since the expansion (\ref{II}) is already performed around the corresponding solution. 
Thus, the effective Lagrangian becomes
\begin{eqnarray}
\label{VIII}
\mathcal{L}^{eff}=\frac{{\rm Tr}\ln\left(\beta(S^{-1}_0+\Sigma_\pi)\right)}{\beta V}
-\mathcal{U}_{MF}
+\langle\overline{q}q\rangle\Sigma_{MF}+\frac{\vec\pi^2}{4G_{PS}}
\end{eqnarray}
with $S_0^{-1}=i\slashed\partial-m^*$ and $\Sigma_\pi=-i\gamma_5\vec\tau\cdot\vec\pi$ being the quark propagator in the quasiparticle approximation and self-energy caused by the pseudoscalar correlations. 
We treat these correlations within the Gaussian approximation (see e.g. Ref. \cite{Blaschke2014} for details). 
For this we expand the trace of the logarithm in Eq. (\ref{VIII}) up to the second order in $\Sigma_\pi$. 
The zeroth order term contributes to the mean-field thermodynamic potential
\begin{eqnarray}
\label{IX}
\Omega_{MF}=-12\int\frac{d{\bf k}}{(2\pi)^3}\left[\epsilon-2T\ln(1-f)\right]+\mathcal{U}_{MF}
-\langle\overline{q}q\rangle\Sigma_{MF},
\end{eqnarray}
where $\epsilon=\sqrt{k^2+m^{*2}}$ and $f=[e^{\beta \epsilon}+1]^{-1}$ are the single particle energy and distribution function of quarks. 
The first order term vanishes due to the tracelessness of the Pauli matrices, while the quadratic term yields the one-loop polarization operator of pions with the four momentum $p$,
\begin{eqnarray}
\label{X}
\Pi_\pi=-\frac{\rm Tr(i\gamma_5 S_0)^2}{\beta V}=
-\frac{\langle\overline{q}q\rangle}{m^*}+12p^2\int\frac{d{\bf k}}{(2\pi)^3}\frac{2f-1}{\epsilon(p^2-4\epsilon^2)}.
\end{eqnarray}
The poles of the propagator $D_\pi^{-1}=\frac{1}{2G_{PS}}-\Pi_\pi$ define the dispersion relation of the pseudoscalar mode, which at zero momentum gives the pion mass $M_\pi$. 
The present model being a generalization of the NJL one provides validity of the Gell-Mann-Oakes-Renner relation
$M_\pi^2F_\pi^2=-m\langle\overline{q}q\rangle$
with the pion decay constant
\begin{eqnarray}
\label{XI}
F_\pi^2=12m^*(m^*-m)\int\frac{d{\bf k}}{(2\pi)^3}\frac{2f-1}{\epsilon(M_\pi^2-4\epsilon^2)}.
\end{eqnarray}
The temperature dependence of $M_\pi$ is shown in the left panel of Fig. \ref{fig1}. 
At low $T$ the pion mass remains almost constant and is significantly smaller than the two quark decay threshold $2m^*$. 
The increase of temperature leads to a rapid decrease of the effective quark mass around the pseudocritical temperature. Above this temperature $M_\pi>2m^*$ and the pion is not a bound state but a resonance. 
Above a certain temperature the real part of $D_\pi$ does not have zeros, which terminates the pion mass curve.

The polar representation of $D_\pi=|D_\pi| e^{i\delta_\pi}$ defines the phase shift of pions $\delta_\pi$. It gives a direct access to the spectral function of pions $\rho_\pi=\frac{1}{\pi}\frac{\partial\delta_\pi}{\partial p_0}$. On the other hand, the field $\vec \pi$ enters $\mathcal{L}_{eff}$ quadratically through the term $\frac{\vec\pi D_\pi \vec\pi}{2}$. By integrating over the fields  $\vec \pi$ we arrive to the thermodynamic potential 
\begin{eqnarray}
\label{XII}
\Omega=\Omega_{MF}-\frac{3{\rm Tr}\ln(\beta^2 D_\pi)}{2\beta V}=\Omega_{MF}+3T\int\frac{d{\bf p}}{(2\pi)^3}\int\frac{dp_0}{\pi}
\ln(1-e^{\beta p_0})\frac{\partial\delta_\pi}{\partial p_0},
\end{eqnarray}
where in the second step the pion contribution is given by the Beth-Uhlenbeck formula. 
The extremum of $\Omega$ with respect to $\langle\overline{q}q\rangle$ defines the chiral condensate. 
The system pressure can be found as $p=-\Omega+\Omega_0$, where the constant shift of energy $\Omega_0$ is introduced in order to ensure $p=0$ in the vacuum. 
The right panel of Fig. \ref{fig1} shows the total pressure as a function of temperature, compared to partial contributions of quarks and pions. 
At low $T$ the pressure of pions dominates the total one and is very close to the pressure of an ideal gas of particles with the mass $M_\pi=140$ MeV. 
At $T\simeq T_c$ the quark pressure exhibits a fast growth while the pion pressure decreases due to the Mott dissociation. 
At high $T$ quarks dominate the thermodynamics of the system.

\section{Color superconducting quark matter at zero temperature}
\label{sec4}

At zero temperature the beyond-mean-field bosonic correlations of quarks can be neglected. 
Therefore, in this section we treat the model within the mean-field approximation. 
In this case all the fields introduced in Eq. (\ref{IV}) are set to some constant values. 
For scalar and pseudoscalar fields $\sigma=\vec\pi=0$. 
By a proper Lorentz transformation only the zeroth component of the vector field attains a finite value $\omega$, which is absorbed into the effective chemical potential of quarks $\mu^*_f=\mu_f+\omega$. 
Furthermore, there exists a global gauge transformation, which makes $\Delta_2^{}$ the only non-vanishing diquark field. Only its absolute value $\Delta=|\langle\Delta_2^{}\rangle|=|\langle\Delta_2^*\rangle|$ enters the thermodynamic potential
\begin{eqnarray}
\label{XIII}
\Omega=-\frac{{\rm Tr}\ln(\beta\mathcal{S}^{-1})}{2\beta V}+\mathcal{U}_{MF}-\langle\overline{q}q\rangle\Sigma_{MF}-
\frac{\omega^2}{4G_V}+\frac{\Delta^2}{4G_D}.
\end{eqnarray}
Solving the trace of the logarithm of the inverse Nambu-Gorkov propagator is a technically demanding but straightforward procedure exhaustively described in the literature (see e.g. Refs. \cite{Baym2018,Blaschke2014}). 
We omit it for brevity the description of our approach. 
Since working within the mean-field approximation, we require chiral condensate, vector and diquark fields to minimize $\Omega$. 
This gives us a set of three coupled equations to be solved with respect to $\langle\overline{q}q\rangle$, $\omega$ and $\Delta$. Having the solutions found, we obtain pressure, baryonic charge and energy density using standard thermodynamic identities as $p=-\Omega+\Omega_0$, $n_B=\frac{\partial p}{\partial\mu_B}$ and $\varepsilon=\mu_B n_B-p$, respectively.

\begin{figure}[t]
\centering
\includegraphics[width=6.4cm,clip]{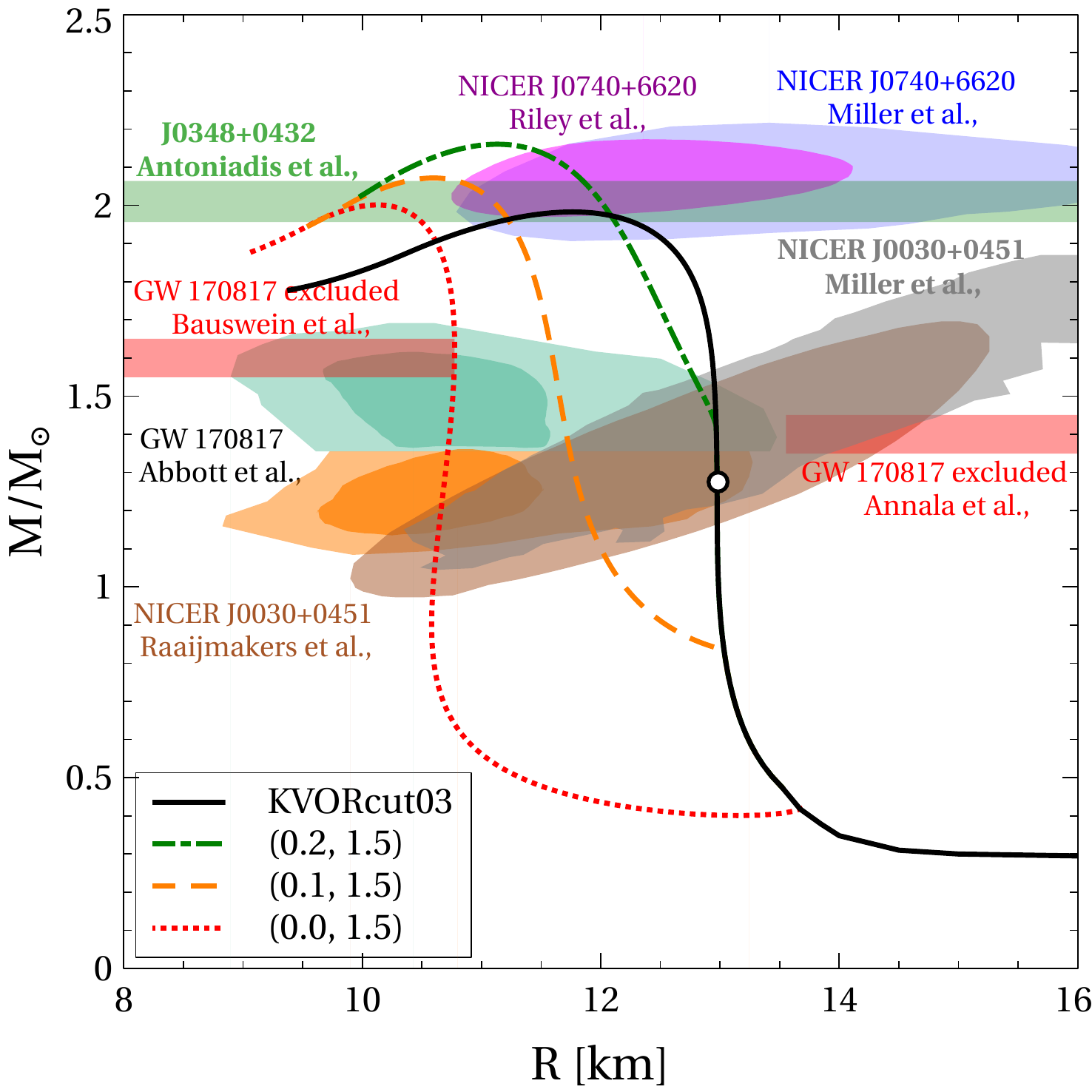}\hspace*{.2cm}
\includegraphics[width=6.4cm,clip]{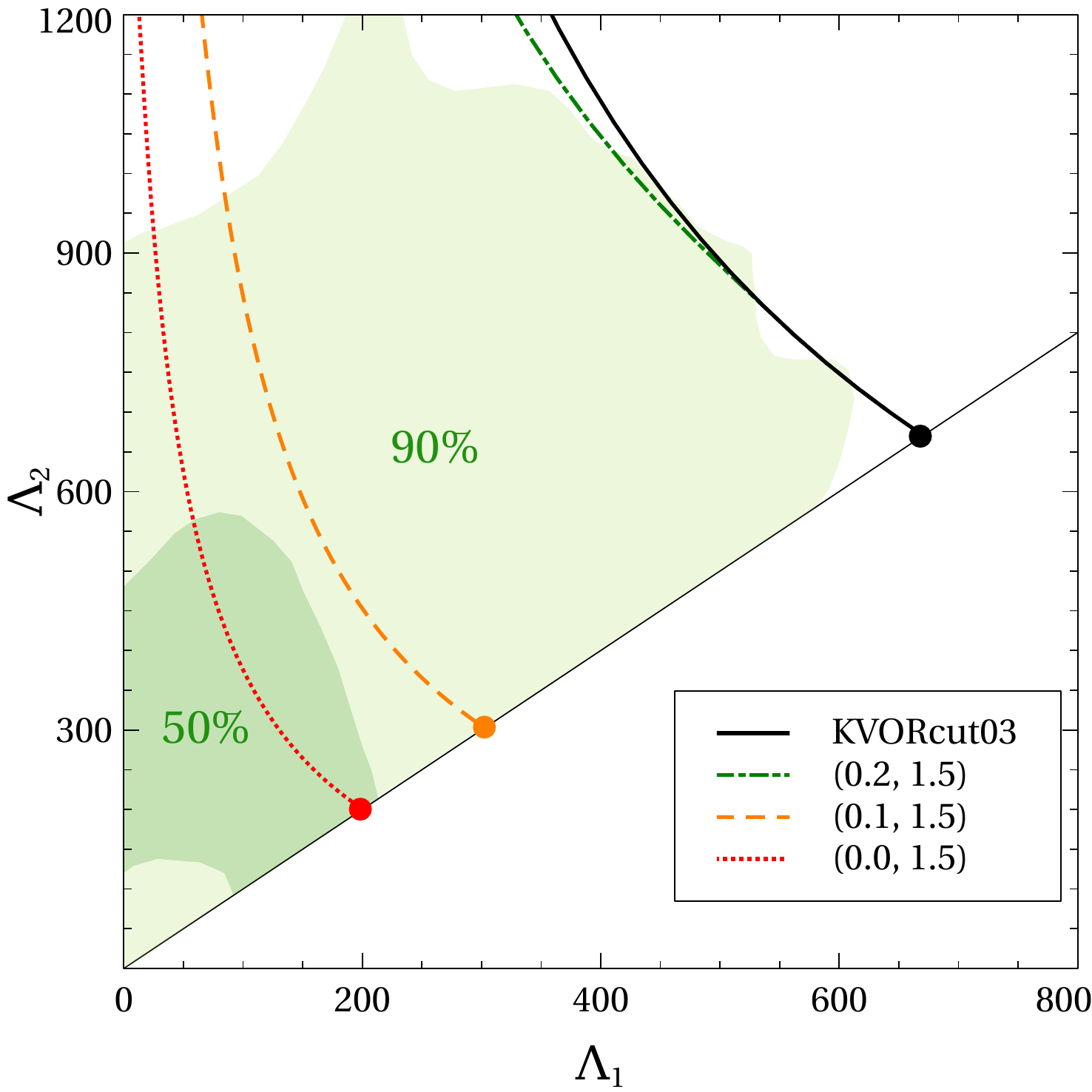}
\caption{{\it Left panel:} mass-radius diagram calculated for the hybrid quark-hadron EoS discussed in the text (colored curves) and purely hadronic DD2npY - T EoS. Empty circle on the hadronic curve indicates the hyperon onset.
The astrophysical constraints depicted by the colored bends and shaded areas are discussed in the text. {\it Right panel:} Tidal deformabilities of two components of the NS-NS merger with the chirp mass $\mathcal{M}=1.188~{\rm M}_\odot$. Colored filled circles represent the configurations with equal masses of two components $\rm M_1=M_2=1.3646~ M_\odot$. Dark and light green shaded ares represent the regions falling into the $50~\%$ and $90~\%$ confidence levels.}
\label{fig2}
\end{figure}

The equation for the diquark pairing gap has two solutions. The trivial one $\Delta=0$ exists at any value of the chemical potential, while $\Delta\neq0$ appears at the 2SC phase onset. It occurs when two solutions coincide. The condition $\frac{\partial^2\Omega}{\partial\Delta^2}\bigl|_{\Delta=0}=0$ should be solved with respect to the baryonic chemical potential in order to find its critical value $\mu_B^c$ corresponding to the 2SC phase onset. Below $\mu_B^c$ quark matter exists in a normal phase, while above it color superconductivity is energetically favorable. 
This critical value decreases with growing $\eta_D$ since a stronger diquark pairing leads to earlier onset of the 2SC phase. The requirement of the absence of color superconductivity in the vacuum sets an upper limit constraint on the diquark coupling $\eta_D<\eta_D^*$. 
This $\eta_D^*$ can be found from the condition $\frac{\partial^2\Omega}{\partial\Delta^2}\bigl|_{\Delta=\mu_B=0}=0$ combined with the vacuum mas gap equation. 
Thus
\begin{eqnarray}
\label{XIV}
\eta_D^*=\frac{3}{2}\frac{m^*}{m^*-m}\frac{G_{PS}}{G_\infty}=2.54,
\end{eqnarray}
which is about 70 \% larger than the corresponding NJL 
value  \cite{Zablocki2010}. 
This difference is caused by the fact that $\eta_D^*$ is obtained from the analysis of the vacuum state where $G_{PS}\simeq2G_\infty$. 
Therefore, $G_D=\eta_D^* G_\infty$ is about twice larger compared to the maximum value allowed by the NJL model. 
This makes the present approach applicable to describing the regime of strong diquark couplings. 

We construct the hybrid EoS of NS matter by imposing the conditions of $\beta$-equilibrium and electric neutrality \cite{Baym2018} as well as joining our EoS for the quark matter phase to the one for the hadronic phase by means of the Maxwell construction. 
The hadronic phase is described by the KVORcut03 model
\cite{Maslov:2015wba},
which is a relativistic density functional model 
that includes hyperons and has a medium dependent ratio of masses and couplings being a function of the scalar density. 
Solving the TOV equations we construct the mass-radius diagram of NS with quark cores shown on the left panel of Fig. \ref{fig2}. For the chosen sets of the vector and diquark couplings the model is able to fit the constraints coming from the observation of PSR J0348+0432 \cite{Antoniadis2013}, PSR J0740+6620 \cite{Riley2021,Miller2021} and PSR J0030+0451 \cite{Raaijmakers2019,Miller2019} pulsars as well as the gravitational wave signal from the merger GW170817 \cite{Bauswein2017,Annala2018}. The model also fits the region of the $2\sigma$ confidence level coming from the analysis of the same gravitational wave signal \cite{Abbott2018}. Getting to the $1\sigma$ region requires an adjustment of $\eta_D$, which is beyond the scope of the present contribution. 

We also compare the results of our model for the tidal deformability $\Lambda$ of NS with the constraints for it extracted from the gravitational wave signal of the inspiral phase of the binary neutron star merger event GW170817 \cite{Abbott2018}. For this we calculate $\Lambda_1$ and $\Lambda_2$ of the two components of the binary system with masses $\rm M_1>M_2$ providing the chirp mass  \cite{Peters1963} $\mathcal{M}=1.188~{\rm M}_\odot$. 
The sets of $\eta_V$ and $\eta_D$ consistent with the discussed mass-radius constraints get only to the 90 \% confidence level interval of the constraint on $\Lambda_1$ and $\Lambda_2$. The agreement, however, can be improved by simultaneous adjustment of the vector and diquark couplings.

\section{Conclusions}
\label{sec5}

We have studied a relativistic density functional approach to quark matter, which i) mimics the quark confinement by a rapid growth of the quark self-energy in the confining region, ii) respects chiral symmetry of strong interaction and iii) can be interpreted as a chiral quark model with the density dependent coupling constants. In addition to the vector repulsion channel, we have introduced the diquark pairing not studied before within the density functional approach.

At zero baryon chemical potential we introduce mesonic correlations beyond the mean-field. For this the most important pseudoscalar mode of quark correlations identified with the pion excitations is considered within the Gaussian approximation. 
The corresponding pion contribution to the thermodynamic potential is analyzed within the Beth-Uhlenbeck formalism. Since the Mott dissociation of pions occurs at roughly the same temperature where the effective quark mass experiences 
a rapid decrease, the present approach exhibits 
a switching between the hadron and quark degrees of freedom with increase of the temperature. 

In order to model NS with quark cores we apply the present density functional approach to construct a hybrid EoS of hadronic and color superconducting quark matter with the vector repulsion. The results of this modeling are confronted to various constraints of the NS mass-radius relation and tidal deformability. The approach provides a reasonable agreement with these constraints, however, having a potential for further improvement by adjustment of the values of the vector and diquark couplings.

\section*{Acknowledgements}
This work has been supported by the Polish National Science Centre (NCN) under grant No. 2019/33/B/ST9/03059. It was performed within a project that has received funding from the European Union’s Horizon 2020 research and innovation program under grant agreement STRONG – 2020 - No 824093. K. M. acknowledges support by the Foundation for the Advancement of Theoretical Physics and Mathematics ``BASIS'', project No. 17-15-568-1. D.B. was supported by the Russian Foundation for Basic Research under grant No. 18-02-40137 and by the Federal Program "Priority-2030".


\end{document}